# FLASH MICROWAVE PRESSING OF ZIRCONIA

Charles Manière[a], Geuntak Lee[a,b], Elisa Torresani[a], John F. Gerling[c], Vadim V. Yakovlev[d], Darold Martin[e], Eugene A. Olevsky[a,f,*]

(a) Powder Technology Laboratory, San Diego State University, San Diego, USA
(b) Mechanical and Aerospace Engineering, University of California, San Diego, La Jolla, USA
(c) Gerling Consulting, Inc. 1122 Stoddard Avenue, Modesto, California, USA
(d) Department of Mathematical Sciences, Worcester Polytechnic Institute, Worcester, Massachusetts, USA
(e) US Army CCDC AC, Picatinny Arsenal, New Jersey, USA
(f) NanoEngineering, University of California, San Diego, La Jolla, USA

**Abstract**

Microwave Pressing is a promising way to reduce microwave sintering temperatures and stabilize microwave powder materials processing. A multi-physics simulation was conducted of the regulated pressure-assisted microwave cavity. This simulation took into consideration resonance phenomena and the nonlinear temperature-dependent material parameters of zirconia. The intrinsic behaviors of microwave systems and zirconia make the regulation of the microwave pressing difficult. However, the same phenomena can be used to activate flash sintering. Flash microwave sintering uses high electric fields of the resonant microwave profile, the Negative Temperature Behavior (NTC) of zirconia resistivity, and the mechanical pressure applied to the powder via a die compaction configuration. The resulting flash microwave pressing still needs improvement in terms of the processed material structure homogeneity, but it has the capacity to become the fastest sintering treatment as it allows room temperature activation where the total process time only takes a few seconds. In addition, this 10-20s processing technique has shown good potential for improving the transparency of alumina pre-sintered specimens.

---

* Corresponding author: **EO**: Powder Technology Laboratory, San Diego State University, 5500 Campanile Drive, San Diego, CA 92182-1323,
Ph.: (619)-594-2420; Fax: (619)-594-3599, *E-mail address*: eolevsky@mail.sdsu.edu

**Nomenclature**

$C_p$ Heat capacity (J·kg$^{-1}$·K$^{-1}$)

$T$ Temperature (K)

$\kappa$ Thermal conductivity (W·m$^{-1}$·K$^{-1}$)

$Q_e$ Heat source (W·m$^{-3}$)

$\varphi_{rsa}$ Surface to ambient radiative heat flux (W·m$^{-2}$)

$\sigma_s$ Stefan Boltzmann constant (5.67E-8 W·m$^{-2}$K$^{-4}$)

$\epsilon$ Emissivity

$T_{air}$ Air temperature (K)

$\varphi_{csa}$ Convective heat flux (W·m$^{-2}$)

$h_{ia}$ Surface conductivity (W·m$^{-2}$·K$^{-1}$)

$J$ Surface *radiosity* (W·m$^{-2}$)

$G$ Irradiation flux (W·m$^{-2}$)

$n$ Refractive index

$e_b(T)$ Surface radiation produced (W·m$^{-2}$)

$\rho_r$ Reflectivity

$\varphi_{rss}$ Net inward radiative heat flux (W·m$^{-2}$)

$\mu_r$ Complex relative permeability

$\varepsilon_r$ Complex relative permittivity

$\mu_r''$ Relative permeability imaginary part

$\varepsilon_r''$ Relative permittivity imaginary part

$k_0$ Vacuum wave number (rad·m$^{-1}$)

$\sigma$ Electric conductivity (S·m$^{-1}$)

$\varepsilon_0$ Vacuum permittivity (8.854187817...×10$^{-12}$ F·m$^{-1}$)

$\mu_0$ Vacuum permeability (1.2566370614...×10$^{-6}$ T·m/A)

$j$ Complex number

$\omega$ Angular frequency (rad·Hz)

$t$ Time (s)

**$E$** Electric field (V·m$^{-1}$)

**$H$** Magnetic field intensity (A·m$^{-1}$)

$L_{PID}$ PID regulated Cavity length (m)

$e(t)$ Regulated -measured temperature error (K)

$K_p$ PID proportional coefficient

$K_I$ PID integral coefficient
$K_D$ PID derivative coefficient

# 1. Introduction

Using microwave energy is a promising way for heating and sintering of ceramics[1,2]. Employing resonance phenomena, this approach is able to attain very high magnitude of the electric field allowing the heating of dielectrics and semiconductive materials[3]. Compared to conventional furnace operations, where the heating is external, the microwave power can be selectively concentrated in powder ceramic specimens (and adjacent tooling) for a direct and volumetric heating[1,4–7]. In addition, microwave heating is very fast, and numerous studies reported a significant reduction of the microwave sintering temperatures compared to conventional sintering[3,8–11]. Therefore, this sintering approach has high potential for energy savings, fast processing times and improvement of the specimen's mechanical properties[12].

At the same time, microwave sintering is known to have inherent problems of stability[13,14]. The electromagnetic properties of the affected materials evolve with temperature and modify the microwave field distribution and resonance in the cavity[7,15]. When materials with negative temperature electrical resistivity coefficient (NTC) are used, they intrinsically tend to heat up with a thermal runaway often accompanied by resonance phenomena which make the heating very difficult to control[7,16,17]. The NTC behavior along with the specimen's external cooling conditions are responsible for the formation of thermal inhomogeneities such as hot spots[13]. The coupling of these multi-physics phenomena makes the thermal regulation of directly microwave heated materials very challenging.

On the other hand, the thermal runaway associated with the NTC resistivity behavior of materials like zirconia is used in field-assisted sintering to activate an ultra-rapid materials processing technique called "flash sintering"[18–20]. This technique represents a very interesting prospect in the sintering field due to the following potentially positive aspects: reduction of sintering time, reduction of grain growth, control of interface reactions, etc. Compared to resistive flash sintering, "microwave flash sintering" enables a contactless and volumetric transmission of heat through microwave[21–23]. Using high field resonant configurations, it is even possible to initiate flash sintering at room temperature[13] and heat dielectric materials which are normally nearly impossible to heat by the traditional flash sintering process[24,25].

In this work, we will first study the resonance and heating behavior of the developed microwave press cavity via the finite element approach[26]. We will then explore the potential of Flash Microwave Pressing (FMP) of zirconia. Finally, we will show the first potential application of FMP by improving the transparency of alumina specimens.

## 2. Experiment and method

### 2.1. Experimental configuration

The microwave press configuration includes a 2.45GHz microwave generator adapted to a WR340 (86 x 43 mm) rectangular waveguide. This type of waveguide is chosen to enable a $TE_{10}$ microwave propagation mode at 2.45GHz. The microwave generator is a 2kW OEM coupled with a 2-port waveguide isolator (10kW, WR340) which absorbs the reflected power to protect the generator. The microwave equipment also includes a dual directional coupler for microwave measurements and a 3-stub impedance tuner to control the microwave transmission across a 30mm aperture iris[27] which separates the generator's part of the applicator section where the heating press is located. The iris is an essential part of the waveguide that allows the resonance control of the electromagnetic field in the applicator zone for a rapid and efficient heating of the specimen. The applicator section is made of a portion of the waveguide ended by a waveguide adjustable short circuit. A 10mm diameter alumina pressing tooling set is inserted in the applicator and surrounded by a (80% alumina-20% silica) Zircar thermal insulation. The applicator waveguide is designed to allow the 10mm diameter alumina punches to enter the cavity without microwave losses and with a loading/unloading metal window. The internal Zircar insulation and alumina pressing die and punch tooling are designed to allow heating without air convection and radiation cooling. The cooling is allowed only by conduction in the alumina tools surrounded by the Zircar insulation. The CAD model of the complete microwave assembly is shown in Figure 1a. This microwave assembly is mounted on a wheeled table allowing for its insertion into an Instron[TM] mechanical testing device, which applies the pressure and records the force/displacement curves (see the mounting on Instron "press" in Figure 1b).

*2.2. Method*

The objective of this work is to study the microwave pressing of zirconia, in particular, under flash sintering conditions. The described microwave pressing setup enables the microwaves to propagate through the iris into the applicator zone, which is a resonant cavity. In this cavity, it is possible to modify the microwave field intensity by adjusting the cavity length using the movable short circuit. The heating can then be regulated by adjusting the short-circuit length to control the electric field strength and the corresponding heat response in the powder bed[3,28]. The simulation using a movable short circuit will be utilized to see the degree of the thermal homogeneity of Proportional Integral Derivative (PID) controlled heating in the pressing tooling and powder specimen. As discussed below, this approach requires advanced PID measurement/controlling devices and, in their absence, endures high thermal inhomogeneity. Instead, this work focuses on inducing an ultra-rapid heating using constant microwave applied power at different cavity resonance levels. The impact of different electric field intensities, heating rate and pressure was studied for zirconia powder specimens.

After analyzing which experimental conditions activated the flash sintering phenomena in the zirconia powder, a spark plasma pre-sintered (SPS'ed) alumina specimen was placed in the zirconia powder bed to study the potential of the FMP for improving the alumina specimen transparency. Based on the work of Kim *et al*[29] the SPS pre-sintering cycle was 25K/min with heating ramp up to 1000°C, then 2K/min to 1125°C and cooling to room temperature. It was shown that with this short sintering time, a translucent alumina specimen can be obtained.

## 3. Theory and calculations

The governing equations of microwave heating are presented first. The boundary conditions are detailed afterwards.

### 3.1. Microwave heating physics

Microwave heating is simulated through an electromagnetic-thermal model (EMT) which couples the Maxwell's and heat transfer equations[6] COMSOL Multiphysics finite element software was used for these simulations. Maxwell's equations can be written as[6]:

$$\nabla \times (\mu_r^{-1} \nabla \times \boldsymbol{E_r}) = k_0^2 \left(\varepsilon_r - \frac{j\sigma}{\omega \varepsilon_0}\right) \boldsymbol{E_r} \qquad (1)$$

with $\boldsymbol{E_r}$ defined by the harmonic electric field expression $\boldsymbol{E} = \boldsymbol{E_r} exp(j\omega t)$.

The heat transfer part of the EMT model is represented by:

$$\rho C_p \frac{\partial T}{\partial t} + \nabla.(-\kappa \nabla T) = Q_e \qquad (2)$$

$$Q_e = \frac{\omega}{2}(\varepsilon_0 \varepsilon_r'' \boldsymbol{E}^2 + \mu_0 \mu_r'' \boldsymbol{H}^2) \qquad (3)$$

The convection of the air is typically modeled through fluid dynamic equations[30]. However, in the present study an experimental configuration was used where the applicator area is completely filled with Zircar insulation and alumina tools. Air was present only in the areas of the waveguide where no heating was detected. Therefore, the convection was not taken into account, and we assumed continuous insulation at the experimental setup junctions. For these conditions, the computation time was about 10 hours (Intel core i7 3.4GHz, 32MoRAM).

### 3.2. Boundary conditions

The external boundary of the 86 mm insulation box surfaces is subjected to a convective flux loss (different from the previously discussed internal convection) and to the surface-to-ambient thermal radiation described by equations:

$$\varphi_{csa} = h_{ia}(T_{air} - T) \qquad (4)$$

$$\varphi_{rsa} = \sigma_s \epsilon (T_{air}^4 - T^4) \qquad (5).$$

The convection coefficient $h_{ia}$ and emissivity $\epsilon$ of 5W·m$^{-2}$·K$^{-1}$ and 0.85, respectively, are defined for the insulation box surfaces[30]. The internal solid/air interfaces are subjected to surface-to-surface thermal radiation. The solid bodies are assumed to be opaque to thermal radiation and the air is assumed to be transparent. The relationship between the incoming thermal irradiation G and the thermal power radiated $e_b(T)$, is defined through the radiosity $J$ expression (total outgoing thermal radiative flux):

$$J = \rho_r G + \epsilon e_b(T) = \rho_r G + \epsilon n^2 \sigma_s T^4 \tag{6}$$

With the ideal gray body simplification, we determined the relationship between the emissivity and reflectivity to be:

$$\epsilon = 1 - \rho_r \tag{7}$$

Then, we obtained the expression of the net inward radiative heat flux $\varphi_{rss}$:

$$\varphi_{rss} = \epsilon(G - e_b(T)) \tag{8}$$

The electromagnetic boundary conditions included perfectly reflective metallic wall surfaces and a TE$_{10}$ port at the generator location in the rectangular waveguide. The TE$_{10}$ port condition simulates the forward microwaves and the absorption of the reflected microwaves. This port condition approximates the microwave generator and the waveguide circulator. A simulation using a PID on the moving short circuit wall is carried out to study the thermal stability of microwave pressing under typical 50K/min constant heating rate. The 50K/min heating rate is imposed by resonance control adjusting the cavity length. This simulation used the following PID expression (9) for the cavity length $L_{PID}$:

$$L_{PID} = K_p e(t) + K_I \int_0^t e(t)d\tau + K_D \frac{de(t)}{dt} \tag{9}$$

A moving mesh using the Arbitrary Lagrangian Eulerian (ALE) method is employed to allow the PID to modify the waveguide length at the short circuit. The material properties are detailed in Ref[15].

## 4. Results

In this section the simulation of the microwave field distribution in the cavity, dependent on the dimensional and thermal parameters, is presented first. Results are then presented of the simulation of the microwave heating through a typical 50K/min heating rate regime. After discussing the microwave and heating behavior of the cavity, the experimental investigations of the microwave pressing are presented in the following order: (i) the model/cavity calibration using the thermal response of a silicon carbide susceptor, (ii) the study of the flash microwave pressing of zirconia, and (iii) the potential application of FMP for improving the transparency of alumina specimens.

### *4.1. Microwave simulation of the waveguide press setup*

In order to study the microwave field distribution and identify the resonant characteristics of the cavity, a parametric steady state simulation is conducted using the cavity dimensions and the tooling temperatures. The steady-state simulation allows testing of a parameter (iris aperture, temperature) with a scan of the waveguide dimensions to locate the resonance profile. The resonance profile corresponds to a maximum of the electric field intensity in the cavity and a minimum of the wave reflection parameter (see an empty applicator in Figure 2).

#### *4.1.1. Effect of the iris aperture*

The resonant profiles corresponding to different iris apertures are plotted in Figure 3. For each of these curves, the maximum of the electric field indicates the cavity length corresponding to the resonance. Increasing the opening size in the iris aperture has two main consequences: (i) shifting of the resonance position to a shorter cavity length, and (ii) broader resonance peaks. This shifting means that the distribution of the field across the iris induces a contraction of the apparent $TE_{10}$ wave length. However, the most interesting aspect is the width of the resonance peak. An overly narrow resonance peak, like in the case of a small iris aperture, is difficult to control. The resonance peak for an aperture of 30mm is more stable while allowing a large increase of the electric field magnitude. Therefore, an aperture of 30mm is chosen.

*4.1.2. Effect of the temperatures and PID regulation strategy*

The dielectric properties of the zirconia powder, alumina tools and Zircar insulator include the complex permittivity, whose real and imaginary parts increase with temperature. At the same time, the apparent wave length is contracted with this increase modifying the resonant properties of the cavity. Understanding this phenomenon is essential for developing an efficient temperature regulation system that is not affected by the resonance change during heating. A parametric simulation was carried out considering three different temperatures. The results are reported in Figure 4. The phenomenon of the contraction of the standing wave is characterized by the shifting of the resonance peak to the shorter cavity length for increasing temperatures. Based on this, it was deduced that the best control strategy was to start from a longer cavity length that was outside of the resonance range (low magnitude of the E field) and impose the PID process control on the short circuit position and move towards the reduction of the cavity length (toward high magnitude of the E field). In this way, the PID reduces the cavity length increasing the E field until the resonance required to heat the powder is obtained. As the resonance position in the cavity is shifted during heating, the controller will follow this shift and reduce the cavity length as needed to maintain the resonance required to heat the powder. The least desirable control strategy would be to start from the left side of the resonance peak in Figure 4 and to move toward resonance by increasing the cavity length. In this configuration, high initial sample temperatures would occur resulting in an uncontrolled thermal runaway. The other thermal phenomena indicated in Figure 4 show that increasing the cavity length, the maximum E field magnitude was reduced, the peak broadened and sample temperatures increased. The reduction of the E field intensity at high temperatures is not a problem because the field coupling at high temperatures is significantly increased and the broader peak stabilizes the control process.

*4.2. Simulation of resonance regulated microwave heating*

In order to test the microwave press heating response, a typical 50K/min heating rate was imposed via a PID controller that adjusted the cavity length (see details in previous section). The simulation results are reported in Figure 5 (a simulation video is provided in supplementary material). Figure 5a shows the area where the ALE moving mesh is applied to the PID in-situ modifying the cavity length. This area can be contracted to increase the electric field and extended to reduce it. Figure 5b shows the intended sample temperature and the measured sample temperature perfectly fitting the thermal cycle (except a small delay in the beginning). An interesting aspect is the regulated cavity length (Figure 5c) and the corresponding electric and temperature fields (Figure 5d and Figure 5e). In the first 50s, a low E field magnitude (high cavity length) is quickly displaced by a high magnitude (closer to resonance) allowing the initial heating of zirconia powder which has limited microwave coupling at room temperature. Figure 6 shows the E field distribution at 250s (high field magnitude) and the contraction of the standing wave close to the sample. At about 300°C the coupling of the zirconia powder with microwave field is higher. Under constant microwave power, the zirconia powder experiences a thermal runaway that started at that temperature[13]. In the present case, the PID controller broke this thermal runaway by increasing the cavity length, which goes back to the lower E field far from resonance. This way, the 50K/min ramp was preserved and the runaway temperature range starts with the abrupt change in the cavity length (Figure 5c). This length change is possibly due to a highly reactive PID controller which was difficult to implement experimentally. Regarding the temperature field, a hotspot is clearly shown during heating (Figure 5e). This hotspot phenomenon originated from the intense cooling of the edges of the powder bed via the surrounding alumina tooling. This tooling represents a significant material mass which takes time to be heated. The hotspot formation is then intrinsic and generated the high thermal instability despite the slow heating rate of 50K/min. In contrast, in the flash sintering configurations (next section), it is expected that extremely fast heating (few seconds) may delay the hotspot formation by reducing the edge cooling flux time.

### 4.3. Experimental calibration of the cavity heating response using silicon carbide

The first experimental investigation was the calibration of the cavity based on the respective simulation. The heating response of the silicon carbide susceptor was tested for different cavity lengths in positions far and close to the resonance predicted by the simulation. Only low temperatures (200-500°C) were explored for this first investigation. The results are reported in Figure 7. The predicted resonance corresponded to a cavity length of 214 mm. The thermal response of the susceptor showed increasing heating rates for positions closer to the predicted resonance. At the same time, the heating was so fast that the susceptor was quasi-instantly destroyed. These first experiments seem to confirm the simulated resonance profile.

### 4.4. Flash microwave pressing of zirconia

Like the previous calibration tests, the FMP experiments were first started by observing the sintering response (displacement curve) for different cavity length positions and for an initial applied pressure of 5MPa. The displacement curve recorded by the Instron mechanical testing device was used as an indicator of sintering advancement. For each flash sintering experiment, the pressure was applied by ramping to full pressure in 1min, and the microwave field was then turned on. The microwaves were turned off when the flash sintering displacement curves reached a plateau indicating sintering had concluded. The FMP displacement curves at 5MPa are reported in Figure 8a. These curves show that the closer the cavity length is to the resonance position (~199mm), the shorter the flash sintering onset time will be. At 204.5mm the total microwave processing time was 500s, for 202.8mm it was 50s and the flash sintering itself took 12s. This microwave flash sintering process, starting at room temperature, has good potential to have an extremely short processing time. The next tests were performed at 10MPa using an intermediate position of 203.5mm (see Figure 8b). The sintering response was similar to the equivalent test at 5MPa. All the specimens subjected to FMP have weakly sintered corner edges, which seem to indicate the presence of a hot spot in the center of all the specimens (like in Figure 5e). Attempts were made to alleviate this problem by adding a 1mm thick sheet of Zircar insulation to the upper, lower and radial edges of the powder bed. The cavity length was increased slightly to 204mm in order to compensate for the increased thermal confinement to avoid runaway heating. The resulting FMP was indeed faster than the one at 203.5mm validating the thermal confinement effect due to the presence of the insulation layer. However, the edges of the specimen were still weakly sintered indicating that the insulation layer was not sufficient to homogenize the heating in the powder bed. Because these insulation layers were highly porous materials, the pressure applied to the powder and to the layer may have reduced their efficiency. Hybrid heating[6] using susceptors able to resist thermal shock may solve this problem.

*4.5. FMP for improving alumina transparency*

Before finding the solution to the intrinsic homogeneity problems of the FMP, it is possible to use this process for improving the consolidation of pre-sintered samples. This can be demonstrated by the improvement of the optical properties of alumina spark plasma pre-sintered samples.

In the domain of transparent ceramics, specimens obtained by SPS often suffer from carbon pollution and reducing atmospheres caused by long sintering time at low temperature and high pressure[31–35]. Based on the work of Kim *et al*[29], translucent alumina specimens were pre-sintered by SPS at 1125°C with no hold time. Then, instead of adding a long holding time at this temperature in the SPS to obtain better transparency, we placed the obtained specimens in the flash microwave press. The translucent alumina specimen was surrounded by the zirconia powder (as seen in Figure 8) for the activation of the flash phenomenon and as a pressure transmitting medium. The results are reported in Figure 9. In the first FMP experiment, which took 50s of sintering time, an improvement of the specimen's transparency at the edge was noted and a highly damaged area in the center of the specimen was seen which was most likely caused by an intense hotspot. In a second FMP experiment, the flash sintering time was reduced to 18s and damage in the center of the sample was avoided while enabling an improvement of the specimen's transparency. A transverse crack was still present in the specimen's volume, similarly to the previous experiment.

FMP allows improvement of translucent alumina into a semi-transparent material. This result seems to validate the recent work of Kerbart *et al*[36], where a microwave sinter-forging setup was successfully used to obtain transparent spinel ceramics starting from pre-sintered specimens. Improvements to the process stated in this paper will be required to increase the transparency of obtained specimens; however, these results are very promising when the 18s microwave treatment is compared to the hours of holding time required by SPS[37,38] (which often result in contamination problems).

## 5. Conclusions

A pressure assisted microwave system has been developed and studied by simulations and experiments. This equipment has been designed to be entirely controlled by the adjustment of the cavity dimensions and the inner resonant profile it develops. An electromagnetic-thermal model has been used to understand the complex coupling of different physics, the evolution of their properties with temperature and, in particular, the resonant behavior of the cavity.

Flash microwave pressing has demonstrated good potential for room temperature activated flash materials treatment with a very short total processing time (~200s with a sintering time of about 20s). However, during FMP, the thermal field develops hotspots and it is nearly impossible to homogenize it in direct heating mode due to the cooling in the surrounding tooling. To homogenize the powder temperatures while avoiding hotspots, it is necessary to provide hybrid heating by using susceptors. This approach would require a special study dedicated to the determination of the optimal susceptor configuration able to resist the thermal shock without shielding the electric field in the powder.

Despite the problems with the structure inhomogeneity when using this flash technique, this approach has shown good potential for improving the transparency of alumina pre-sintered specimens obtained by spark plasma sintering. FMP is promising due to its ability to partially collapse the residual porosity within a total processing time as short as 150s (with a pure sintering time of 18s). Further research is needed for achieving better control of the temperature homogeneity to avoid hotspot damage and crack nucleation in the specimens subjected to FMP.

**Acknowledgements**

The support of Minority Serving Institutions Science, Technology, Engineering and Mathematics Research and Development Consortium (MSRDC) (Award No: D01-W911-SR-14-2-0023) is gratefully appreciated. We are grateful to Gerling Applied Engineering, Inc. for providing the microwave generator and waveguides for this study.

**Figure captions**

Figure 1 a) Microwave line assembly from the generator to the applicator, b) Microwave furnace installed in the Instron mechanical testing device.

Figure 2 Typical resonant profile of the microwave press cavity for an empty applicator using a parametric steady state simulation with different waveguide dimensions; the left graph represents the cavity maximum electric field (where the resonance corresponds to the maximum of the peak) and the right graph represents the reflective coefficient S11 (where the resonance corresponds to the minimum of the peak).

Figure 3 Parametric study of the resonant profile for different iris aperture; the cavity length scan is conducted for each aperture with representation of the cavity maximum electric field; insert presents a cross-sectional view of the applicator area showing the iris and the aperture (red arrow).

Figure 4 Parametric study of the resonant profile for different tooling temperatures; the red arrow represents the control strategy including the start from a low resonance and decrease of the cavity length following the resonance evolution with the tooling temperature.

Figure 5 Simulation of the microwave press heating (50K/min); a) scheme of the waveguide with indication of the area affected by the ALE movable mesh controlled by the PID regulation, b) regulated average sample temperature (blue) and set temperature (green circles), c) plot of the total cavity length which is modified to regulate the specimen temperature, d) cavity electric field distribution and e) temperature field.

Figure 6 Electric field and cavity dimensions at 250s (simulation figure 5) when the electric field is maximum during the 50K/min heating cycle.

Figure 7 Experimental calibration of the microwave cavity and the simulation using the heating response of a silicon carbide (SiC) susceptor at different fixed applicator cavity lengths; the simulated electric field at resonance is reported in the lower part.

Figure 8 FMP displacement curves for different fixed application cavity lengths; the microwave field is turned on at the beginning of the pressure plateau (blue curve) and turned off when the sintering displacement plateau is observed; a) experiments at 5MPa, b)

experiments at 10MPa; all obtained flash microwave pressed pellets have fragile edges (see photo below).

Figure 9 Flash microwave pressing of the pre-sintered translucent alumina samples. The 10mm pre-sintered by SPS samples were polished at the edges to fit the inner 9mm microwave die; the samples are surrounded by the zirconia powder for activation of the flash phenomenon; the flash experimental displacement/pressure curves are reported on the right.

**F1**

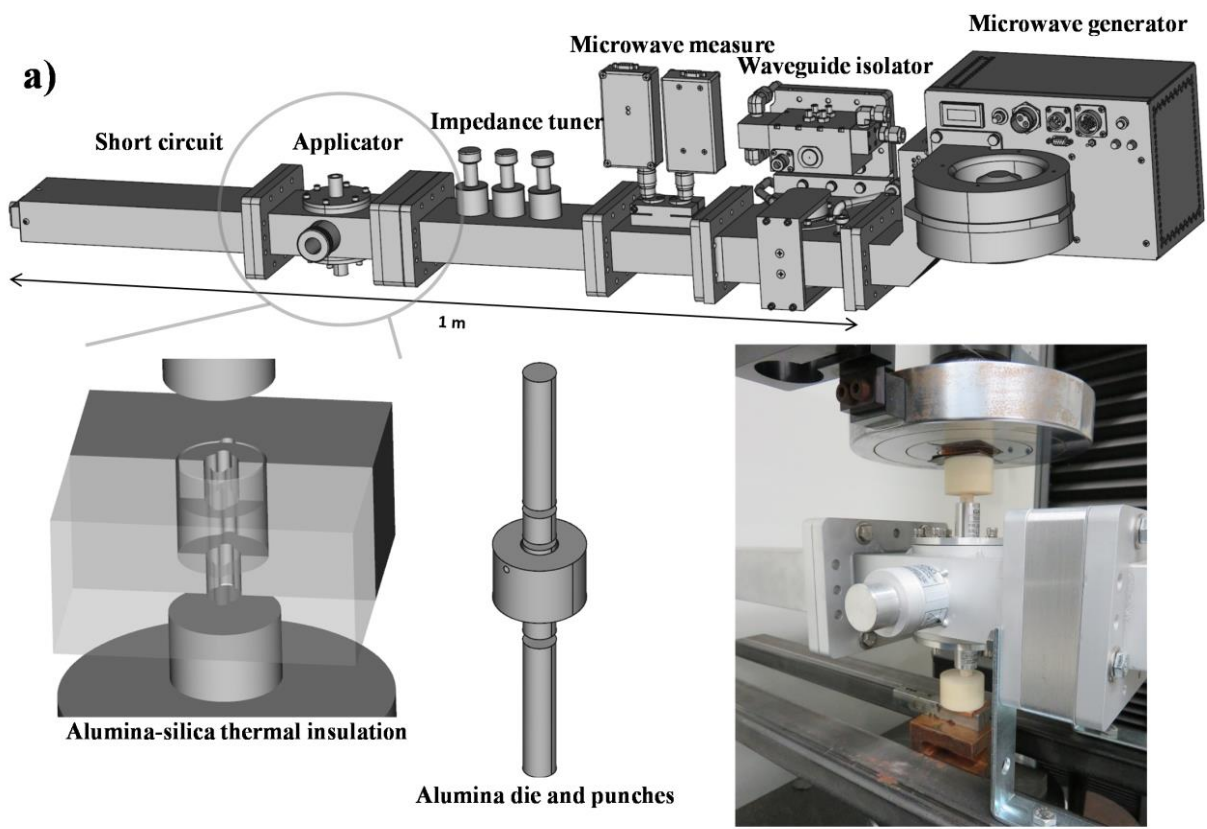

**F2**

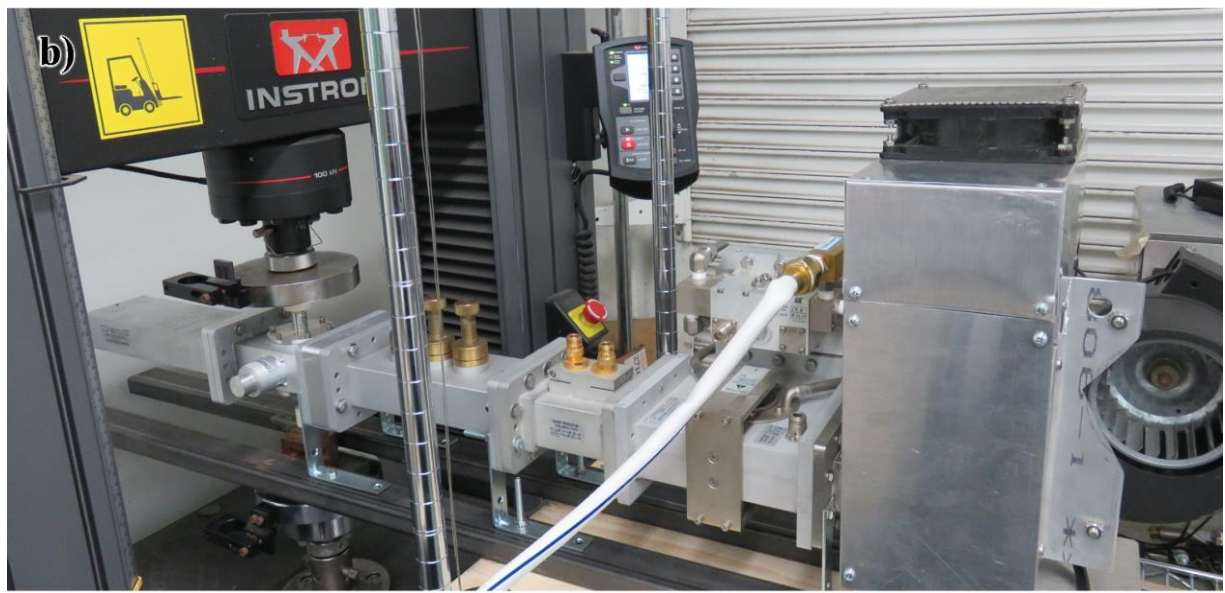

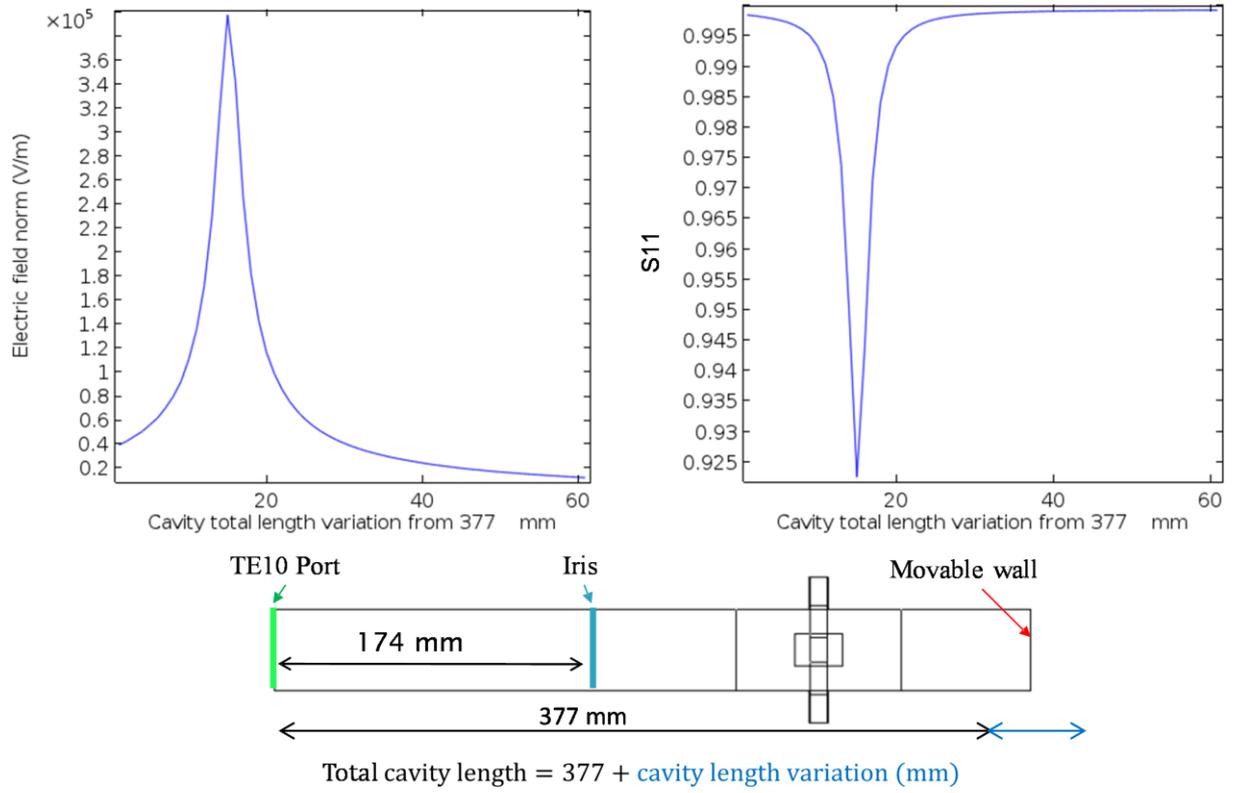

**F3**

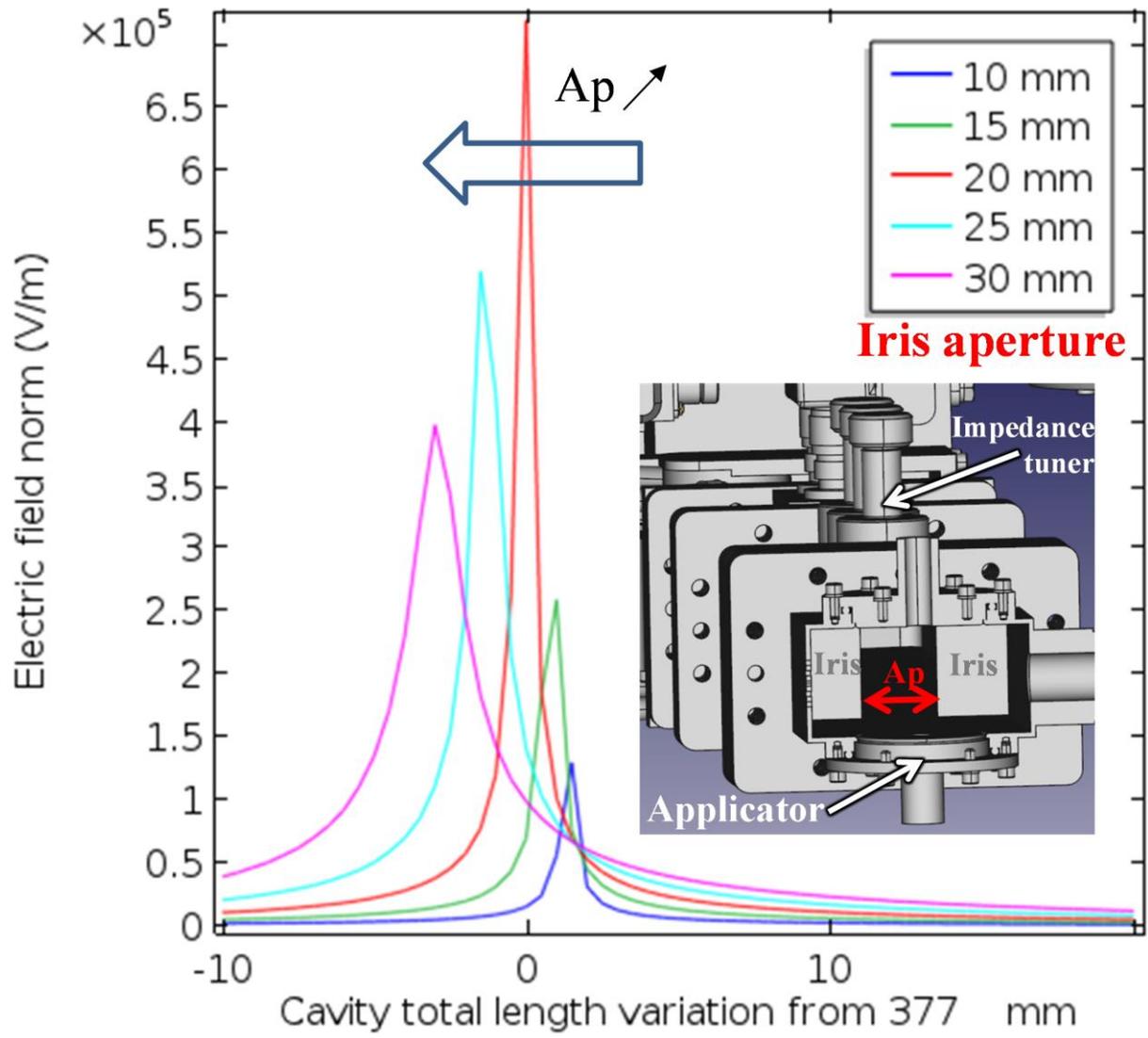

**F4**

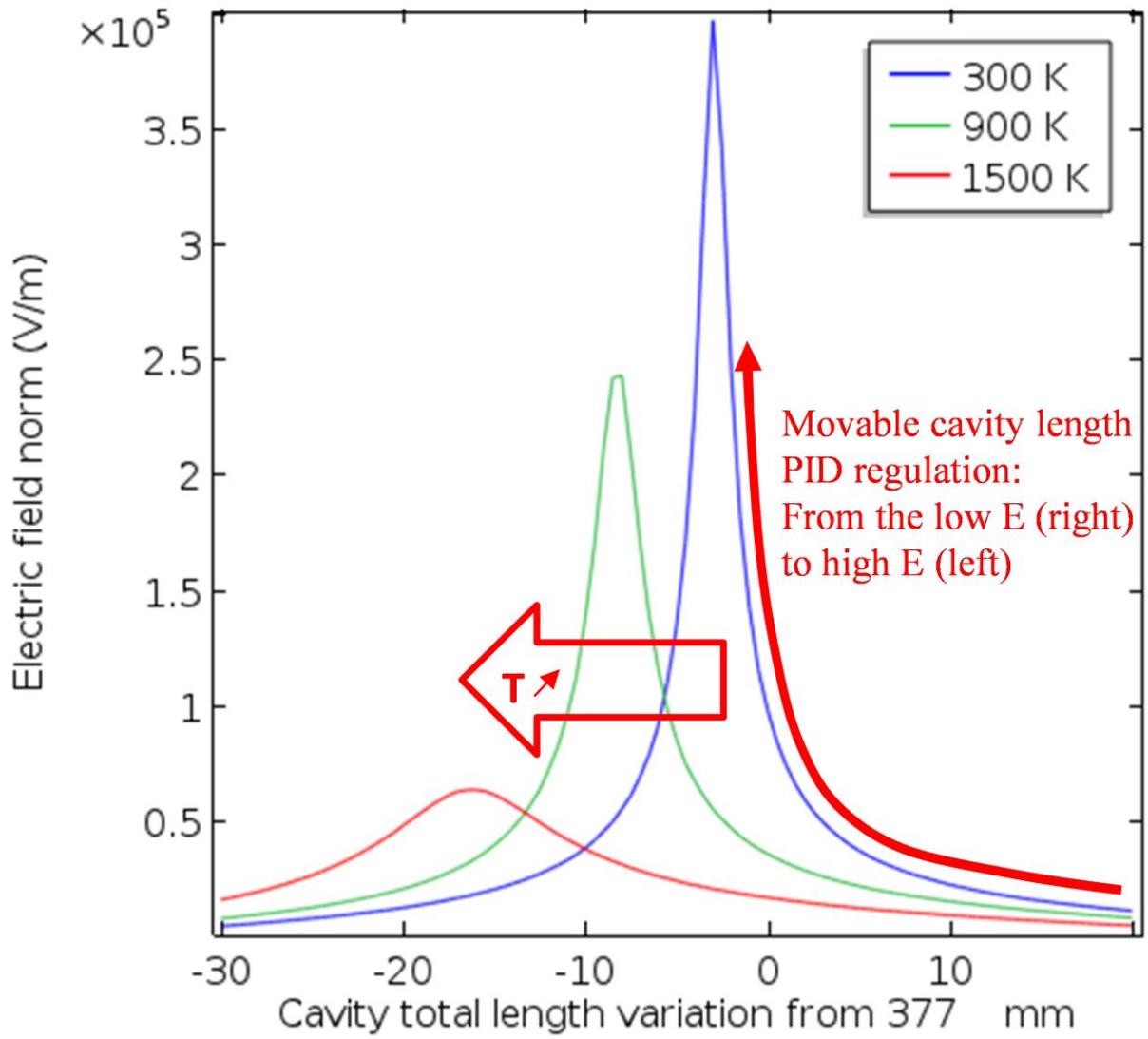

F5

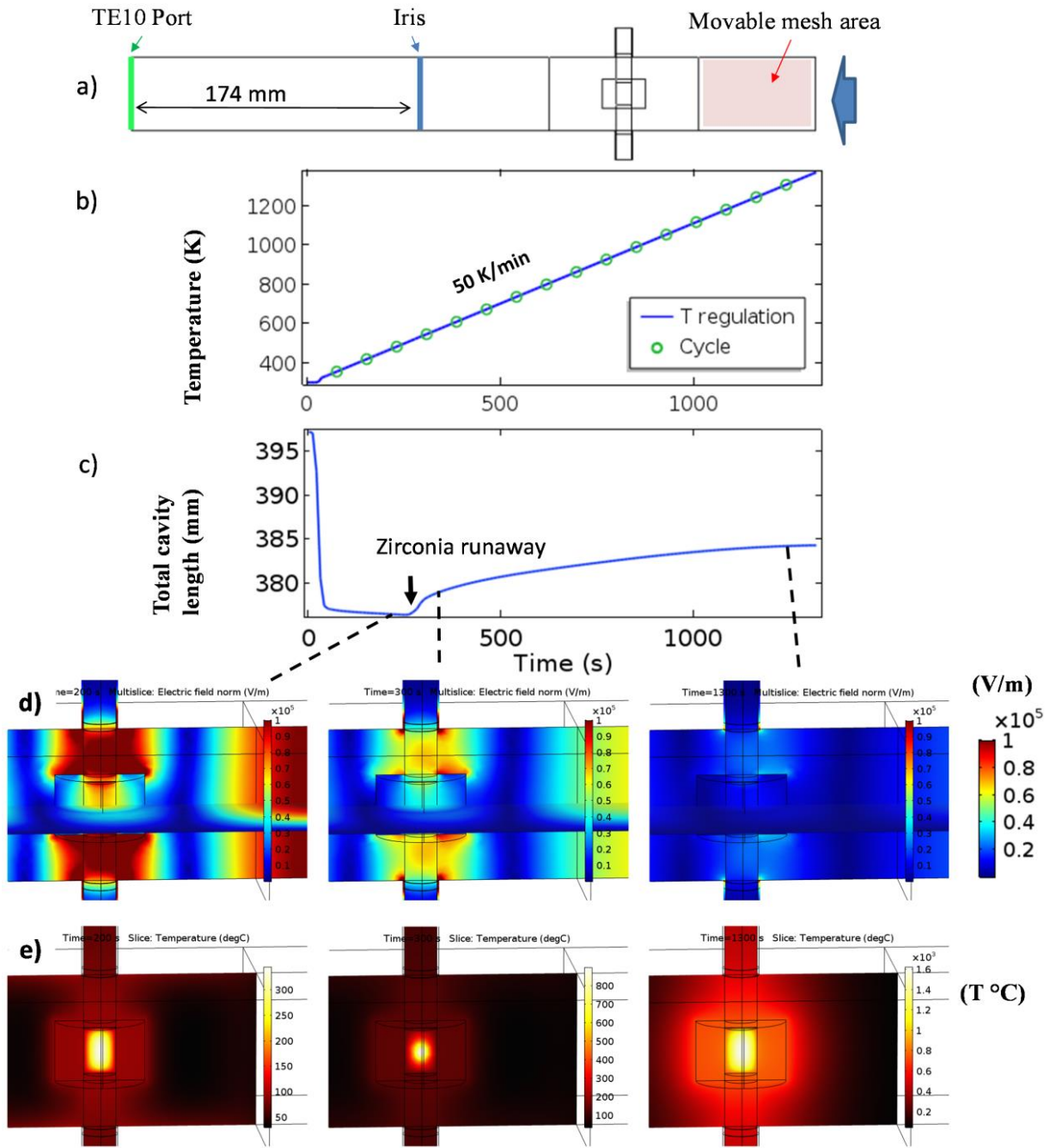

F6

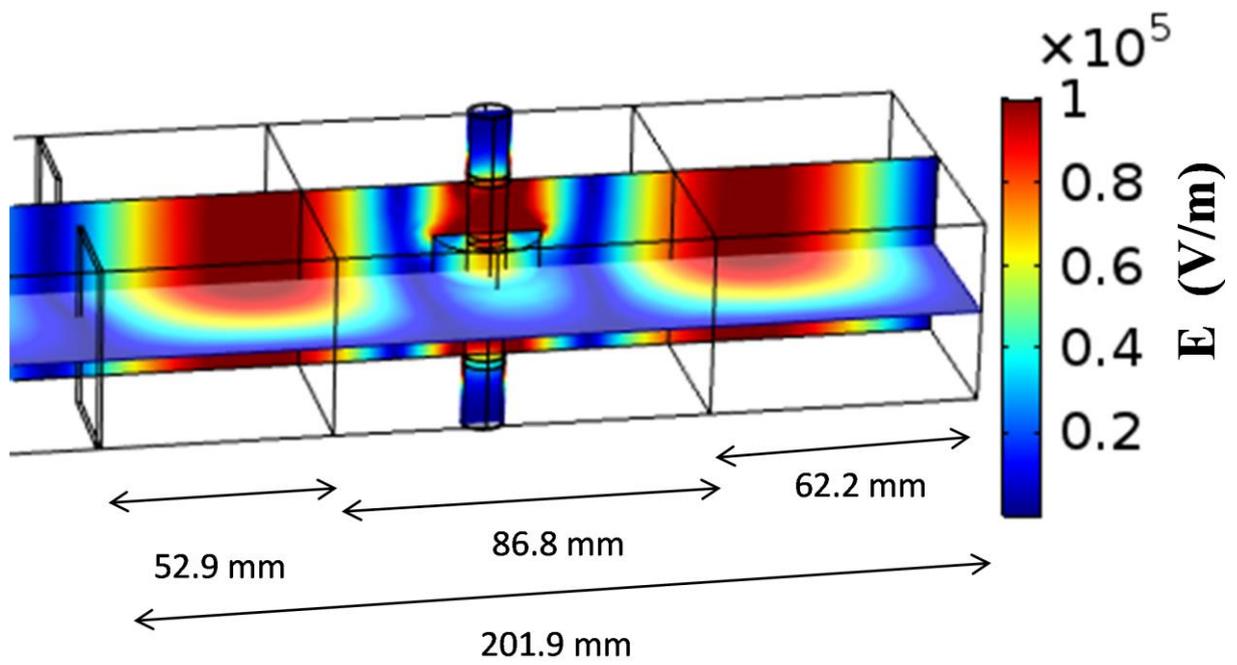

F7

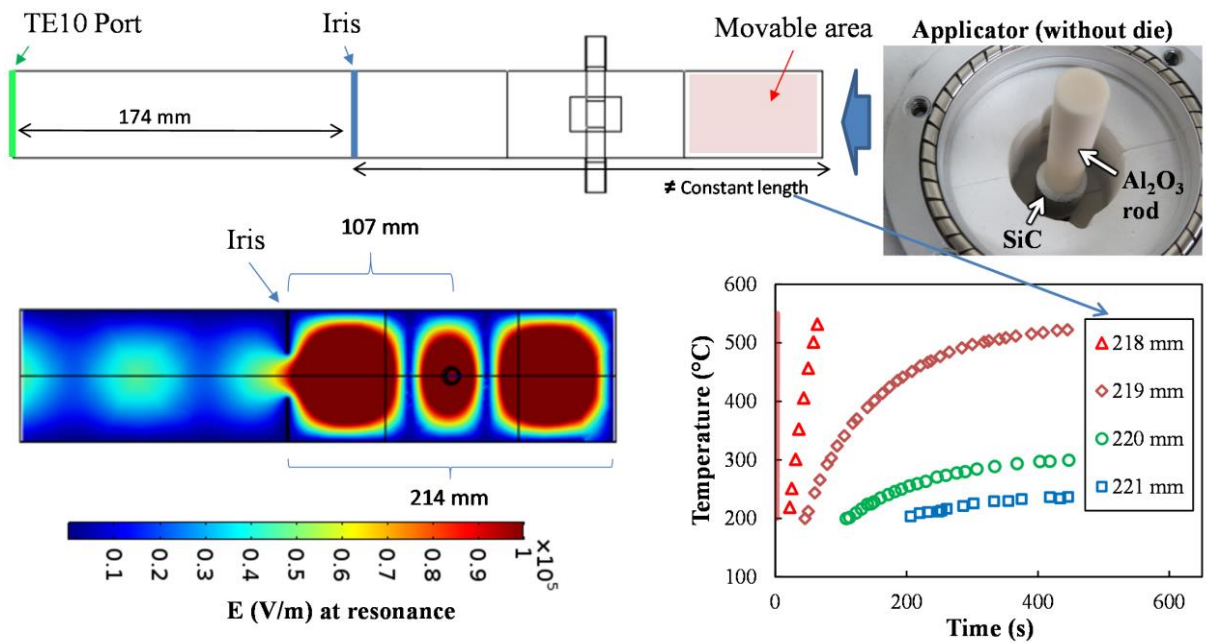

F8

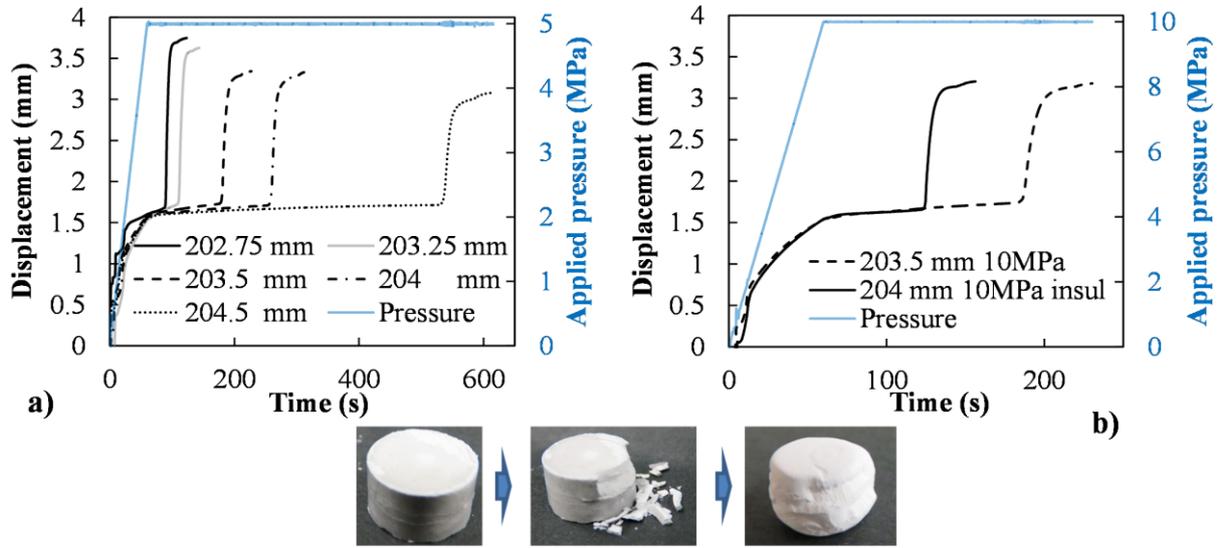

F9

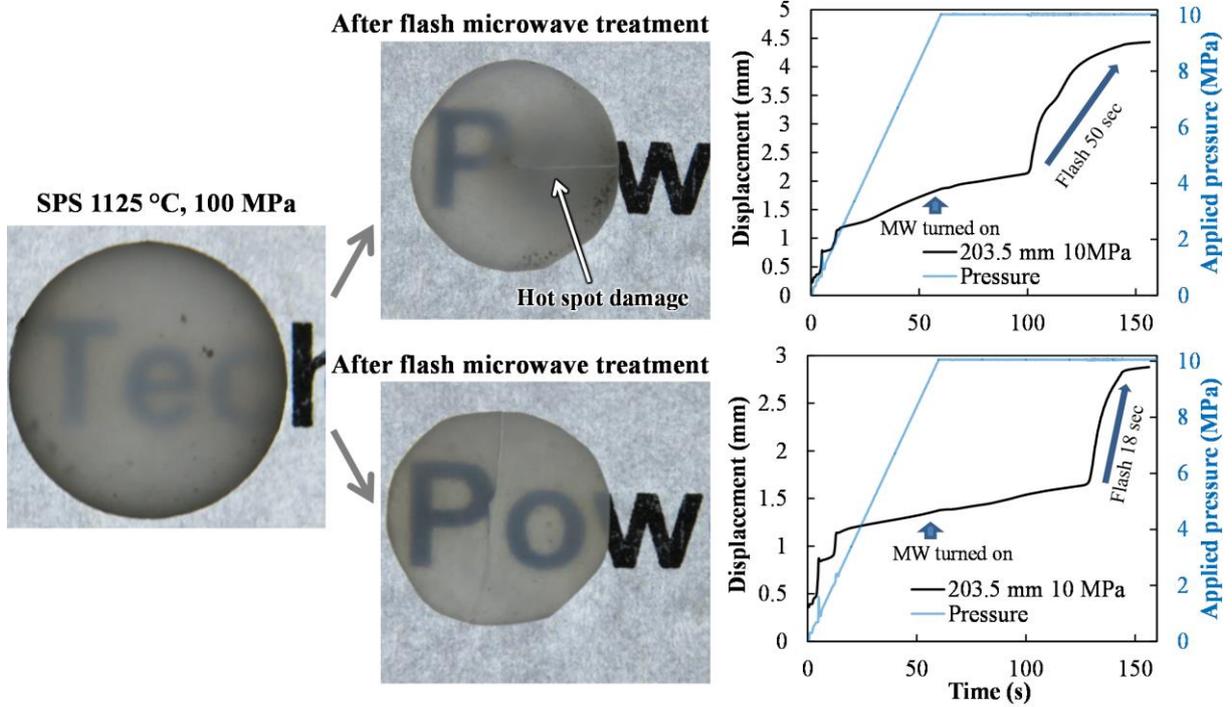